\documentclass[reprint,
aps,prb,twocolumn,superscriptaddress,
floatfix,
showpacs,amsmath,amssymb
]{revtex4-1}

\usepackage{verbatim}
\usepackage{graphicx,color,bbm}
\usepackage{epsfig}

\newcommand{\ket}[1]{|{#1}\rangle  }
\newcommand{\braket}[2]{\langle{#1}| {#2}\rangle}
\newcommand{\ketbra}[2]{\vert {#1} \rangle \langle{#2}\vert}

\begin{document}
\preprint{APS/123-QED}

\title{Population transfer in a Lambda system 
induced by detunings}

\author{P.G. Di Stefano}
\affiliation{Dipartimento di Fisica e Astronomia,
Universit\`a di Catania, Via Santa Sofia 64, 95123 Catania, Italy.}
\author{E. Paladino}
\affiliation{Dipartimento di Fisica e Astronomia,
Universit\`a di Catania, Via Santa Sofia 64, 95123 Catania, Italy.}
\affiliation{CNR-IMM  UOS Universit\`a (MATIS), 
Consiglio Nazionale delle Ricerche, Via Santa Sofia 64, 95123 Catania, Italy.}
\affiliation{Istituto Nazionale di Fisica Nucleare, Via Santa Sofia 64, 95123 Catania, Italy.}
\author{A. D'Arrigo}
\affiliation{Dipartimento di Fisica e Astronomia,
Universit\`a di Catania, Via Santa Sofia 64, 95123 Catania, Italy.}
\affiliation{CNR-IMM  UOS Universit\`a (MATIS), 
Consiglio Nazionale delle Ricerche, Via Santa Sofia 64, 95123 Catania, Italy.}
\author{G. Falci}
\email[gfalci@dmfci.unict.it]{}
\affiliation{Dipartimento di Fisica e Astronomia,
Universit\`a di Catania, Via Santa Sofia 64, 95123 Catania, Italy.}
\affiliation{CNR-IMM  UOS Universit\`a (MATIS), 
Consiglio Nazionale delle Ricerche, Via Santa Sofia 64, 95123 Catania, Italy.}
\affiliation{Istituto Nazionale di Fisica Nucleare, Via Santa Sofia 64, 95123 Catania, Italy.}

\date{\today}
\begin{abstract}
In this paper we propose a new protocol to achieve 
coherent population transfer between two states in a 
three-level atom by using two ac fields. It is based on the 
physics of Stimulated Raman Adiabatic Passage (STIRAP), but 
it is implemented with the constraint of a reduced control, 
namely one of the fields cannot be switched off. A combination
of frequency chirps is used with resonant fields, 
allowing to achieve approximate destructive interference, 
despite of the fact that an exact dark state does not exist. 
This new chirped STIRAP protocol
is tailored for applications to artificial atoms, where 
architectures with several elementary units can be  
strongly coupled but where the possibility of switching 
on and off such couplings is often very limited. 
Demonstration of this protocol would be a benchmark for the 
implementation of a class of multilevel advanced control 
procedures for quantum computation and microwave quantum 
photonics in artificial atoms.
\end{abstract}

\maketitle
\section{Introduction}
Preparation of a quantum system in a well defined state is 
an essential task in many branches of modern physics 
ranging from atomic and molecular 
physics~\cite{kr:201-vitanov-annurev} 
to quantum computation~\cite{kb:210-nielsenchuang}.
Techniques for transferring population from a ground 
state $\ket{0}$ to a state $\ket{1}$ employ either 
Rabi cycling or adiabatic 
passage (AP)~\cite{kr:201-vitanov-advmolopt}. 
Amongst these latter 
STIRAP is a three-level atom scheme where selective and 
faithful population transfer is achieved by operating with 
two resonant driving fields in $\Lambda$ 
configuration~\cite{kr:198-bergmann-rmp-stirap,kr:201-vitanov-advmolopt}.
The advantage over Rabi cycling is the dramatic reduction 
of sensitivity to fluctuations of the parameters, 
at the expenses of a longer duration 
of the adiabatic protocol. In more complex architectures 
semiclassical driving fields are substituted by harmonic modes 
of a strongly coupled cavity, and tasks as preparation of
photons with controlled amplitude, frequency and 
polarization~\cite{ka:202-kuhnetal-prl-singleph,ka:213-muckerempe-pra-stirapsinglephgen} can be performed by 
vacuum-stimulated Raman AP (vSTIRAP).

In the last few years multilevel coherence 
in solid-state systems, from mesoscopic 
devices~\cite{kr:211-younori-nature-multilevel} to atomic-like
impurity states~\cite{ka:208-kleinhalfmann-pra-stirapdoped}, 
has been a fertile subject of investigation.
In particular superconducting-based 
``artificial atoms''~\cite{kr:208-clarkewilhelm-nature-squbit,kr:211-nori-repprogphys-artatoms,kr:213-devoretschoelkopf-science} are very
promising since coherent phenomena proper of the microscopic 
realm have been demonstrated on the mesoscopic 
scale.  
With respect to their natural counterpart, artificial atoms 
offer the advantage that composite structures can be 
fabricated on a single 
chip~\cite{kr:208-schoelkopf-nature-wiring}, 
which allows manipulation of single photons 
at $\mathrm{GHz}$ frequencies opening the new scenario of 
microwave quantum 
photonics~\cite{ka:213-nakamurayam-ieee-microwphot}.
Moreover new architectures could be implemented with both larger 
couplings~\cite{ka:210-niemczyck-natphys-ultrastrong} and 
a larger degree of integration than their atomic 
counterparts.

In the last few years 
several theoretical proposals~\cite{ka:204-muraliorlando-prl-eit,ka:205-liunori-prl-adiabaticpassage,ka:206-siebrafalci-optcomm-stirap,ka:209-siebrafalci-prb,ka:208-weinori-prl-stirapqcomp,ka:212-falci-physscr,ka:213-falci-prb-stirapcpb} 
and experiments~\cite{ka:209-sillanpasimmonds-prl-autlertownes,ka:209-baurwallraff-prl-autlertownes,ka:212-lihakonen-srep-qswitch,ka:210-kellypappas-prl-cpt} have dealt with multilevel 
coherence in artificial atoms.  Distinctive 
features of such systems are the effectiveness of 
parity selection rules~\cite{ka:205-liunori-prl-adiabaticpassage,ka:209-siebrafalci-prb,ka:213-falci-prb-stirapcpb} which 
together with the presence of strong $1/f$ 
noise~\cite{ka:205-falci-prl,kr:214-paladino-rmp}, impose 
constraints on the available control. 
Therefore new protocols for manipulating the 
coherent dynamics must be tailored for such systems.  
Their design requires that large couplings allowing for efficient control are combined with protection 
from noise~\cite{ka:213-falci-prb-stirapcpb}.

In this paper we present a new protocol to achieve coherent 
population transfer between the two lowest excited states of 
a three-level atom by using two ac fields. 
The key difference with standard STIRAP, where ac fields must be 
switched on and off in a counterintuitive sequence~\cite{kr:198-bergmann-rmp-stirap}, is that one of the fields is kept always-on, 
its amplitude being constant during the protocol. 
Operations require phase modulation, and for this reason we
call the protocol cSTIRAP (chirped STIRAP). 
Sweeping the frequency 
of a single classical driving field across 
the resonance is a standard way to switch on and off 
Rabi oscillations, thereby one may think to rephrase STIRAP 
accordingly, but this is not the case. 
Indeed coherent population requires 
destructive interference of the two 
fields~\cite{ka:209-cohentannoudji-kosmos}. This is
guaranteed by cSTIRAP, which thereby solves a non-trivial 
control problem, its experimental demonstration in artificial 
atoms being by itself an important proof of principle 
of advanced three-level control. Even more interestingly, 
cSTIRAP could apply to architectures where ``artificial atoms'' 
are coupled to 
quantized modes, electromagnetic or nanomechanical, 
where strong coupling is achieved by non-switchable hardware 
elements keeping the interaction always-on.
The protocol we propose possesses certain advantageous 
distinctive characteristics: 
(i) it works with reduced available control, as 
always-on field, (ii) it operates with nearly resonant fields,
reducing the operation time; (iii) it may rely on better 
techniques to control the phase of microwave circuits, 
(iv) it is cyclic.   

The paper is organized as follows. 
In Sec.\ref{sec:co-po-tra} we introduce the model Hamiltonian 
and briefly review standard implementations 
of coherent population transfer in two and three-level atoms. 
In Sec.~\ref{sec:cSTIRAP-mech} we illustrate the new protocol 
discussing in Sec.~\ref{sec:cSTIRAP-sens} the robustness 
against parametric fluctuations and in Sec.~\ref{sec:decoherence}
decoherence effects. Finally, in 
Sec.\ref{sec:conclusions}, along with the conclusions, 
we will discuss the comparison of cSTIRAP with 
other protocols for population transfer operated by frequency 
chirps.

\section{Coherent population transfer in Lambda atoms}
\label{sec:co-po-tra}
In two-level systems
coherent population transfer $\ket{0} \to \ket{1}$ by AP 
is performed by shining a direct coupling field 
whose detuning is swept 
through the resonance at the Bohr frequency of the 
transition. Common examples are Rapid AP (RAP) 
or Stark Chirped RAP (SCRAP)~\cite{kr:201-vitanov-annurev}.

In three-level systems population transfer 
may be achieved in absence of direct coupling, via  
a third \textit{linkage} state $\ket{2}$, coupled 
to  $\ket{0}$ and  $\ket{1}$ by 
a pump field at frequency $\omega_p \simeq E_2-E_0$ and a 
Stokes field at $\omega_s \simeq E_2-E_1$, respectively. 
In particular the {\em Lambda configuration} depicted in the top 
inset of Fig.~\ref{fig:lambda} will be considered in this 
work. Since $\ket{2}$ is usually short-lived, one of 
the goals of coherent techniques is to use 
$\ket{2}$ but {\em never populate it}.
This is achieved in a very efficient and elegant way 
relying on destructive 
interference~\cite{ka:209-cohentannoudji-kosmos}.
The Hamiltonian in the rotating wave 
approximation, in the basis of the bare states 
$\{\ket{0},\ket{1},\ket{2}\}$, is expressed 
in a doubly rotating frame as 
\begin{equation}
\label{eq:H}
H = \begin{bmatrix}
0 								& 	0 							& \frac{1}{2}\Omega^\ast_p(t)	\\
0 								& 	\delta (t)					& \frac{1}{2}\Omega^\ast_s(t) 	\\
\frac{1}{2}\Omega_p(t) 		& 	\frac{1}{2}\Omega_s(t) 	& \delta_p(t) 						\\
\end{bmatrix}
\end{equation}
where $\Omega_k(t)$ with $k=p,s$ are the Rabi frequencies 
of the pump and the Stokes fields, which are 
detuned by $\delta_p := E_2-E_0-\omega_p$ and 
$\delta_s := E_2-E_1-\omega_s$ respectively. 
A key quantity is the two-photon detuning, 
defined as $\delta := \delta_p - \delta_s$. 

Conventional STIRAP relies on the fact that 
at two-photon resonance, $\delta(t) = 0$,
an instantaneous eigenvector 
with zero eigenvalue $\epsilon_0=0$ exists  
given by 
\begin{equation}
\label{eq:dark-state}
\ket{D(t)} = \frac{\Omega_s(t) \ket{0} - \Omega_p(t) \ket{1}}{\sqrt{\Omega_s^2(t) + \Omega_p^2(t)}}
\end{equation}
It is called \textit{dark state}
since population is confined in the 
``trapped subspace'' $\{\ket{0},\ket{1}\}$,  
despite of the fact that the two fields excite both the $0\to2$ 
and the $1\to2$ transitions. 
The key phenomenon preventing population of $\ket{2}$ 
is destructive interference between the amplitudes corresponding 
to the two absorption 
patterns~\cite{kr:201-vitanov-annurev,ka:209-cohentannoudji-kosmos}. Conventional STIRAP consists 
in letting the dark state evolve adiabatically from 
$\ket{D(-\infty)}=\ket{0}$ to 
$\ket{D(\infty)}=\ket{1}$. 
This is achieved by shining pulses 
$\Omega_k(t)$ in a ``counterintuitive'' 
sequence, the Stokes at first and then the pump. An important 
characteristic of STIRAP is the fact that AP is operated when 
both fields are on, determining a two-photon effective coupling
$\ket{0}\leftrightarrow \ket{1}$. 
STIRAP has been observed in a variety of physical  
systems~\cite{kr:198-bergmann-rmp-stirap,kr:201-vitanov-advmolopt}. 
The two-photon character of population transfer, and the 
fact that the protocol is maximally efficient with fully 
resonant fields, $\delta = \delta_s=\delta_p=0$ is the key for 
interesting applications with quantized fields.   

Another three-level technique, Raman Chirped Adiabatic Passage 
(RCAP)~\cite{ka:199-solabergmann-pra-rcap}, 
uses instead phase modulation. 
Population transfer is achieved by two far off-resonance 
chirped laser pulse sweeping through resonance 
(see Sec.~\ref{sec:conclusions}). 
Unlike conventional STIRAP, two-photon resonance is not 
kept during the whole process, causing a transient population 
of state $\ket{2}$ to appear. The latter in principle can be 
made small by accurate tuning of parameters.

\section{Coherent population transfer with an always-on field}
\label{sec:cSTIRAP-mech}
In this section we will address the problem of achieving 
$\ket{0}\to\ket{1}$ population transfer subject to 
two constraints, namely (a) 
keeping the population of $\ket{2}$ small and (b) 
operating with a reduced control, in particular with one of the
fields, for instance the Stokes one, kept always on,
$\Omega_s(t)=:\Omega_0\neq 0$. 
Naively one could suppose that 
sweeping the detuning $\delta_s(t)$ could allow to 
effectively switch on and off $\Omega_s$, allowing again for  
conventional STIRAP. However this is not the case because 
coherent population transfer requires 
that the two-photon resonance condition, $\delta =0 $, 
is kept while sweeping $\delta_s(t)$, to ensure 
destructive interference. 
In what follows we will seek for 
a solution allowing to achieve complete 
population transfer by properly shaping the detunings. 

\begin{figure}[t!]
\centering
\includegraphics[width=0.8\columnwidth]{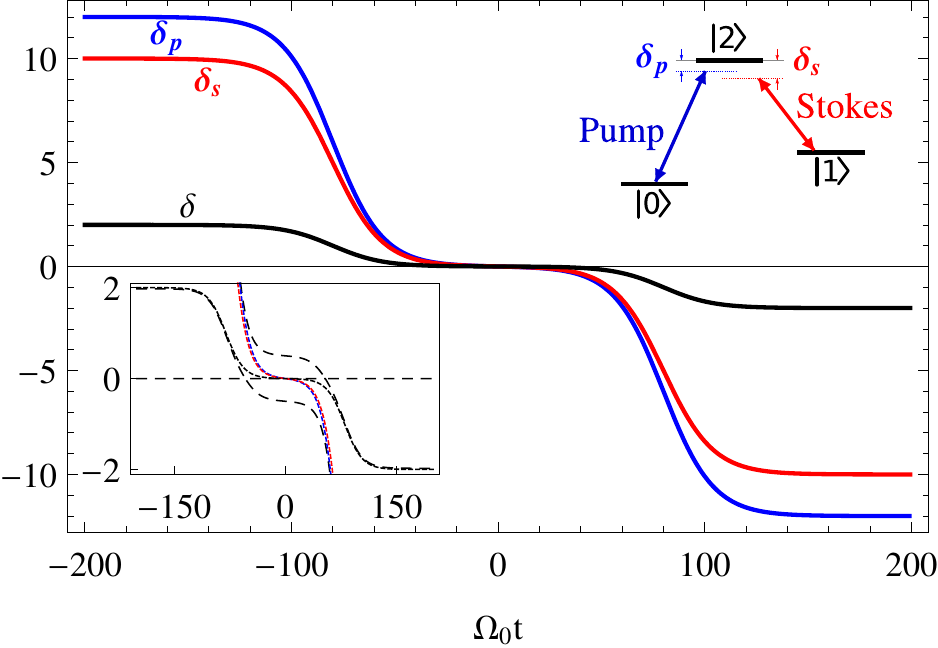}
\caption{(color online) Main figure: single (coloured lines) and two-photon (dotted line) detunings in $\Omega_0$ units. Top inset: Three level Lambda system. Bottom inset: Zoom of the single and two-photon detunings (solid lines), plotted together with the Stokes eigenvalues (dashed lines) of Eq. (\ref{eq:diabatic-eigenvectors}) showing the appearance of a dynamical Stokes-induced AT,
which is switched on and off by modulation of $\delta_s$.
\label{fig:lambda}
}
\end{figure}

First of all when one of the fields is always on, 
the Hamiltonian (\ref{eq:H}) for $t \to \pm \infty$ is not 
diagonal in the bare state basis. In order to approximate 
asymptotically the desired target state $\ket{1}$, necessarily 
at the end of the protocol
we must have $\delta_s \gg \Omega_0$. If we take 
detunings shaped as shown in Fig.\ref{fig:lambda}, 
which are given by
\begin{equation}
\label{eq:detunings}
\begin{aligned}
\delta_s(t) &= \frac{1}{2} \, 
h_{\delta} \Omega_0 \left[\tanh\left(\frac{t-\tau}{\tau_{ch}}\right) + \tanh\left(\frac{t+\tau}{\tau_{ch}}\right)\right] \\
\delta_p(t) &= \kappa_{\delta} \delta_s(t)
\end{aligned}
\end{equation}
the desired asymptotics is ensured by 
$h_\delta \gg 1$, i.e. the protocol must start 
and end with ``far detuned'' lasers. 
The modulation (\ref{eq:detunings}) has the important 
characteristics that at least for part of the protocol 
$\delta(t)=0$ (Fig.~\ref{fig:lambda}). 
During this phase a Stokes-induced Autler-Townes 
(AT) splitting opens. Although an 
exact adiabatic dark state is not 
available for population transfer, we will argue later that 
keeping $\delta \approx 0$ allows to minimize 
the transient population of $\ket{2}$. 

The population transfer mechanism is better understood studying 
the evolution of the instantaneous eigenvalues and eigenvectors 
of the ``Stokes'' Hamiltonian, obtained setting to zero
the pump field in Eq.(\ref{eq:H})
\begin{equation}
\label{eq:Hs}
H_s(t) = 
\begin{bmatrix}
0 								& 	0 							& 0	\\
0 								& 	\delta (t)					& \frac{1}{2}\Omega_0 	\\
0 		& 	\frac{1}{2}\Omega_0 	& \delta_p(t) 						\\
\end{bmatrix}
\end{equation}
Here the Rabi frequency has been taken real 
with no loss of generality.   
The Stokes Hamiltonian acts non-trivially only on the 
$\{\ket{1},\ket{2}\}$ subspace, yielding the asymptotic states
\begin{equation}
\begin{aligned}
\ket{s_+(-\infty)} &\simeq \ket{2}\to\ket{s_+(+\infty)} \simeq \ket{1}
\\
\ket{s_-(-\infty)} &\simeq \ket{1}\to\ket{s_-(+\infty)} \simeq \ket{2}
\end{aligned}
\end{equation}
The "Stokes eigenvalues" display 
the presence of the AT splitting during the protocol 
(Fig.~\ref{fig:lambda}, bottom inset)
\begin{equation}
\label{eq:diabatic-eigenvectors}
s_0 = 0, \quad
s_{\pm} = \delta + \frac{\delta_s \pm \sqrt{\delta_s^2 + \Omega_0^2}}{2}
\end{equation}
During this AT phase $\delta_s$ is swept across the resonance
swapping $\ket{1}\leftrightarrow\ket{2}$. 

\begin{figure}[t!]
\centering
\includegraphics[width=0.8\columnwidth]{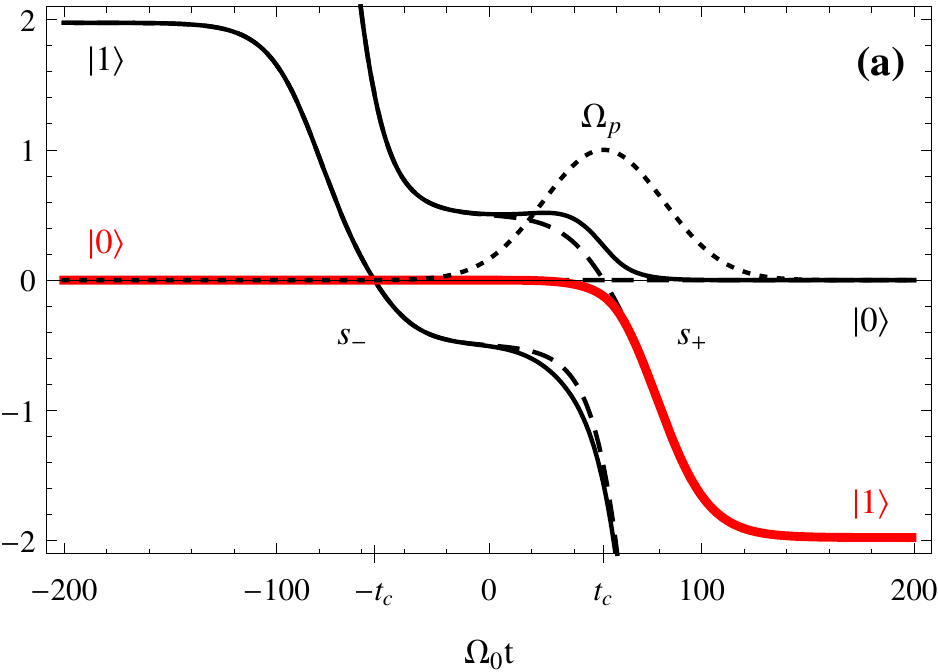}
\includegraphics[width=0.8\columnwidth]{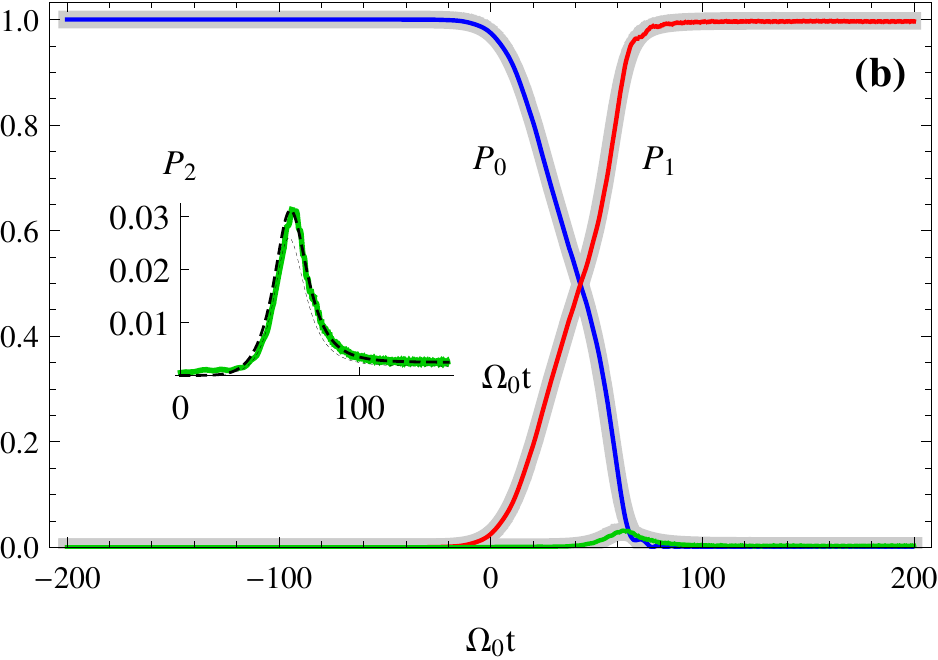}
\caption{(color online) (a) Eigenvalues of the Stokes Hamiltonian of Eq. (\ref{eq:Hs}) (dashed lines) and of the full Hamiltonian of Eq. (\ref{eq:H}) (solid lines) in $\Omega_0$ units. The red thick line is the instantaneous energy of the system adiabatically driven from $\ket{0}$ to $\ket{s_+}\simeq \ket{1}$ through the opening of the avoided crossing generated by the pump pulse (dotted line) at time $t=t_c$. (b) Population histories (red, blue and green lines) from 
the numerical solution of the Schr\"odinger equation, 
for $\Omega_s(t) = \Omega_0, \Omega_0T=40, h_{\delta}=10, \kappa_{\delta}=1.2$ and $\kappa= 1$, $\tau_{ch}=0.6T$, 
showing complete population transfer by cSTIRAP. 
For these parameters the adiabatic 
approximation (gray curves) fully agrees with the exact solution.
Inset: the exact population $P_2$ of the excited state 
(green solid line) is small at any time of the protocol, 
as can be estimated by Eq.(~\ref{eq:P2-adel})
(thin line). The dashed line refers to the approximation of Eq. (\ref{eq:P2-adel-stokes}).
\label{fig:eigenvalues}
}
\end{figure}

Using detunings Eq.(\ref{eq:detunings}) 
with $\kappa_{\delta} > 1$ the pattern of split instantaneous 
eigenvalues $s_\pm(t)$ is crossed twice by the eigenvalue 
$s_0 = 0$, as shown in Fig.~\ref{fig:eigenvalues}(a). 
Crossings occur at times $\pm t_c$ when 
$s_{\pm}(t) = 0$, i.e.   
$4\delta(t_c)\delta_p(t_c)=\Omega_0^2$. 
In these conditions the system prepared in 
$\ket{\psi(-\infty)}=\ket{0}$ remains of course in this state, 
passing through the crossing. Population
transfer is achieved by applying a pump 
pulse with finite area reaching its peak value
close to the second crossing, $t=t_c$. For instance, employing a 
a Gaussian pulse, we have
\begin{equation}
\label{eq:pulse}
\Omega_p(t) = \kappa \,\Omega_0 \mathrm{e}^{-\left(\frac{t-t_c}{T}\right)^2}
\end{equation}
The behavior is understood in terms of the instantaneous 
eigenenergies of the full Hamiltonian Eq. (\ref{eq:H}). 
In particular the pump pulse lifts the degeneracy between 
$s_0$ and $s_+$ turning their crossing into an avoided crossing
[Fig.\ref{fig:eigenvalues}(a)]. The adiabatic connection 
corresponding to $s_+$ yields eventually the desired population 
transfer,
$\ket{0}\to  \ket{s_+(+\infty)}
\simeq\ket{1}$.

We remark that population transfer depends only  
on the presence of a crossing 
between Stokes eigenenergies $s_+$ and $s_0$ and 
on the fact that the adiabatic approximation is valid. 
In this regime the precise shape of the pulses is not 
relevant. Therefore the 
protocol is robust against imperfections in the control. 
From the physical point of view it is worth stressing that 
the pump pulse triggers AP by a two-photon process. 
The distinctive feature of our proposal is that this two-photon effective coupling is 
obtained with both quasi resonant pump and Stokes fields. This 
ensures large efficiency for rather small pulse 
duration. We mention that during its switching on $\Omega_p$ could in principle trigger 
unwanted transitions $\ket{0}\to \ket{2}$, which are 
however suppressed by the Stokes-induced AT splitting and the two-photon resonance condition. 
A similar phenomenon occurs in standard STIRAP, 
where it is called the Stokes-induced EIT (electromagnetically 
induced transparency) phase~\cite{kr:201-vitanov-advmolopt}.

Summing up cSTIRAP can be described in the language of 
Ref.~\onlinecite{kr:201-vitanov-advmolopt} 
as a five stages protocol, with successive far-detuned, 
Stokes-induced AT, Stokes-induced EIT, two-photon AP and again 
far-detuned phases. In what follows we will see that the other 
important requirement, 
namely that  population of $\ket{2}$ should be minimal 
at all times, is also fulfilled. This requirement is necessary 
in order to prevent unwanted decay processes likely to 
occur in real physical systems, where $\ket{2}$ is often 
unstable. 

We estimate 
$P_2 = |\braket{2}{\psi}|^2$ by adiabatic elimination.
The standard procedure formulated in the bare basis yields 
the state $\ket{\psi_{AE}^0}=c_0\ket{0}+c_1\ket{1}$ (see App.~\ref{sec:app-adiabatic}).
First order corrections yield a leakage from the subspace $\{\ket{0},\ket{1}\}$  
given by~\cite{ka:199-solabergmann-pra-rcap}
\begin{equation}
\label{eq:P2-adel}
P_2(t) \simeq \left | \frac{\Omega_p c_0 + \Omega_0 c_1}{2\delta_p} \right |^2
\end{equation}
that can be made very small, as shown in Fig.\ref{fig:eigenvalues}(b), 
which also shows that this approximation works very well. 
A better approximation is obtained by working in the 
Stokes basis [see App.~\ref{sec:app-adiabatic} and Fig.~\ref{fig:eigenvalues}(b)], 
but Eq.(\ref{eq:P2-adel}) has a simpler analytic form, 
allowing to write a figure 
of merit for the parametric dependence of leakage during 
the protocol. 
A simple choice is to consider leakage at 
the crossing $s_+=0$
\begin{equation}
\label{eq:P2-parametric}
P_2(t_c) \simeq \frac{\delta}{\delta_p} f(\kappa) = \frac{\kappa_{\delta}-1}{\kappa_{\delta}} f(\kappa)
\end{equation}
Here $f(\kappa)$ is a monotonically decreasing function of 
the ratio of the Rabi peak amplitudes $\kappa$.
This qualitative behaviour is confirmed by 
numerical simulations shown in 
Fig.~\ref{fig:sensitivity-amplitude}, 
where the efficiency is plotted versus relative magnitude of 
the amplitudes ($\kappa$, left panel) and of the 
detunings ($\kappa_\delta$, right panel), both in the absence 
(top panel) and the in presence (bottom panel) of a finite 
lifetime $\tau_2 = T/2$ of the intermediate state $\ket{2}$ (see section \ref{sec:cSTIRAP-sens} for a model). 
It is seen that efficiency increases with increasing $\kappa$ 
as an effect of a larger avoided crossing at $s_+=0$. 
Moreover increasing $\kappa$ reduces the transient population 
of $\ket{2}$, as given by the figure of merit Eq.(\ref{eq:P2-parametric}).
This is seen by comparing the insets of the left panels of 
Fig.~\ref{fig:sensitivity-amplitude}: the positive slope of the sensitivity in the presence of a finite $\tau_2$ [Fig.~\ref{fig:sensitivity-amplitude}(c)] cannot be explained as an improvement 
in adiabaticity, since this slope is not present in the 
ideal case [Fig.~\ref{fig:sensitivity-amplitude}(a)]. 
Therefore, it can only be caused by a reduction of $P_2$.
Population transfer occurs only for $\kappa_{\delta}>1$ as shown in Fig.~\ref{fig:sensitivity-amplitude}(b),(d). 
In particular, for $\kappa_{\delta} = 1$ we have $\delta(t)=0$ and Eq.(\ref{eq:dark-state}) 
applies, showing that an always on Stokes field 
would produce a return of the population to the initial state. 
For $\kappa_{\delta} < 1$ the Stokes eigenvalues do not cross, 
and adiabatic dynamics leads to a final population entirely 
in $\ket{0}$.

\section{Sensitivity to parameters}
\label{sec:cSTIRAP-sens}

\begin{figure}[t]
\begin{minipage}[b]{0.45\columnwidth}
\centering
\includegraphics[width=\columnwidth]{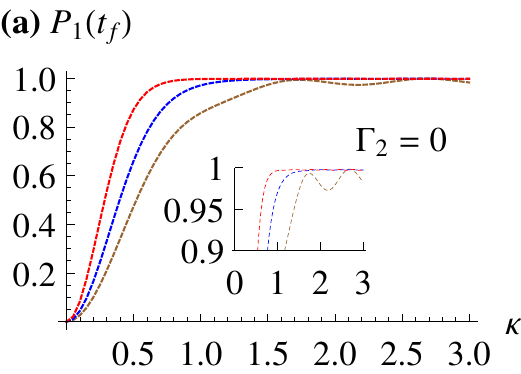}
\end{minipage}
\begin{minipage}[b]{0.45\columnwidth}
\includegraphics[width=\columnwidth]{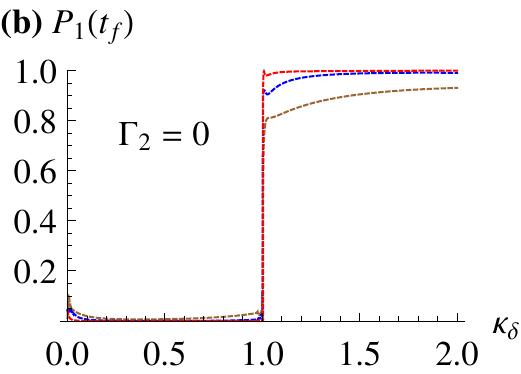}
\end{minipage}
\vfill
\begin{minipage}[b]{0.45\columnwidth}
\centering
\includegraphics[width=\columnwidth]{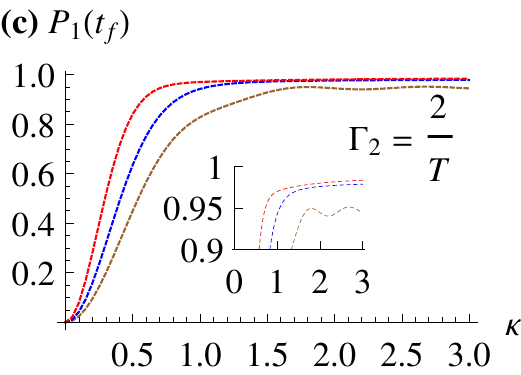}
\end{minipage}
\begin{minipage}[b]{0.45\columnwidth}
\includegraphics[width=\columnwidth]{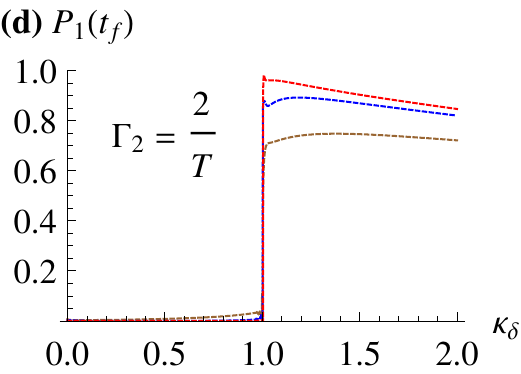}
\end{minipage}
\caption{(color online) Upper panels: STIRAP efficiency vs 
the relative peak amplitudes of the fields 
(left panel, where $\kappa_\delta= 1.2$) 
and to the relative detunings (right panel, where $\kappa=1$), 
for various degrees of adiabaticity (curves: 
$\Omega_0 T=$ 40 (red), 20 (blue), 10 (brown) from higher to lower efficiency). 
For $\kappa_\delta>1$, and 
provided adiabaticity is good, the system has a very slight 
sensitivity to these parameters. 
Lower panels: sensitivity of the efficiency to 
unwanted transient population of $\ket{2}$ 
accounted for by a finite lifetime $\tau_2=:1/\Gamma_2$
(cf. Eq.\ref{eq:parametric-hamiltonian}). The insets of panels 
(a) and (c) are zooms of the corresponding main figures 
showing how, in the presence of a non-vanishing $\Gamma_2$, the efficiency improves with increasing $\kappa$.
\label{fig:sensitivity-amplitude}
}
\end{figure}

The efficiency of cSTIRAP  is not very sensitive 
to slight deviations of relative amplitudes $\kappa$ 
and detunings $\kappa_\delta$ of the pulses, provided adiabaticity is kept. 
This is shown in Fig.~\ref{fig:sensitivity-amplitude}, 
results in the lower panels allowing to fix 
convenient values 
$\kappa=1$, $\kappa_\delta=1.2$ and $\Omega_0 T = 40$, which we 
will use hereafter.

As in conventional 
STIRAP~\cite{kr:201-vitanov-advmolopt}, 
the most critical feature is the parametric sensitivity 
to stray detunings. Here we discuss this issue, which is also 
responsible for decoherence due to low-frequency 
noise~\cite{kr:214-paladino-rmp,ka:213-falci-prb-stirapcpb}.

The physics is understood recalling the picture 
of conventional STIRAP, where two kind of errors 
emerge~\cite{kr:201-vitanov-annurev}. 
``Bad projection'' errors, due a bad choice of the pulse 
shape and timing, may lead to the wrong target state.
``Bad adiabaticity'' errors induce leakage from the 
trapped subspace, nonadiabatic transitions  
surely occurring when the so called 
``global condition'' $\Omega_k T \gg 1$ is not met. 
Both kinds of errors are also triggered by fluctuations induced 
by an environment (see Sec.\ref{sec:decoherence}).
For cSTIRAP we verified that large enough $\Omega_k T$ 
again guarantees adiabaticity (Fig.~\ref{fig:sensitivity-amplitude}). 
In this regime a strong asset of cSTIRAP is that it is not  
affected by bad projection errors in the far-detuned phases, 
since final eigenstates in the rotating frame are nondegenerate.

However since the efficency of cSTIRAP depends on the structure 
of crossings of the eigenvalues of the Stokes Hamiltonian, 
it may be affected by stray detunings during the protocol. 
A further drawback comes from the fact that the state carrying 
population in cSTIRAP, while taking advantage from destructive 
interference, it is not an exact dark state
as in Eq.(\ref{eq:dark-state}), since  
the condition $\delta(t)=0$ does not hold true.
This is a potentially important source of error for cSTIRAP 
since it also may determine a nonvanising population of 
$\ket{2}$ at intermediate times. Sensitivity to detunings 
is conveniently studied by the non-Hermitian Hamiltonian:
\begin{equation}
\label{eq:parametric-hamiltonian}
H(t|\{\delta_k\}) \to H(t|\{\delta_k\}) 
+ i \Gamma_2 \ketbra{2}{2}
\end{equation}
Using a sufficiently large $\Gamma_2 > 1/T$ guarantees 
that transient population of $\ket{2}$ decays elsewhere 
(e.g. in a continuum), yielding a lack of normalization 
at the end of the protocol. Therefore the resulting 
efficiency $P_1(t_f)$ is a figure of merit 
embedding the requirement that $\ket{2}$ should be never 
populated.

The Hamiltonian 
(\ref{eq:parametric-hamiltonian}), where 
only the dependence on detunings is emphasized, 
accounts for the effect of stray components by letting 
\begin{equation}
\label{eq:detunings-fluctuations}
\begin{aligned}&
\delta_k(t) \to \delta_k(t) + \tilde{\delta}_k, \quad k = s,p 
\\&
\delta(t) \to \delta(t) + \tilde{\delta}, \quad  
\tilde{\delta} :=  \tilde{\delta}_p -  \tilde{\delta}_s
\end{aligned}
\end{equation}
Stray detunings may describe very slow 
phase fluctuations (at frequencies $\ll 1/T$) of 
the driving fields. Physically in solid-state devices they 
describe energy fluctuations due
to coupling to an environment 
(see Sec.\ref{sec:decoherence} and Ref.~\onlinecite{ka:213-falci-prb-stirapcpb}) whose power spectrum has $1/f^\alpha$ 
behavior~\cite{kr:214-paladino-rmp}.
In what follows we describe
the detrimental effects they produce and the limitations 
they determine.

\begin{figure}[t!]
\includegraphics[width=0.8\columnwidth]{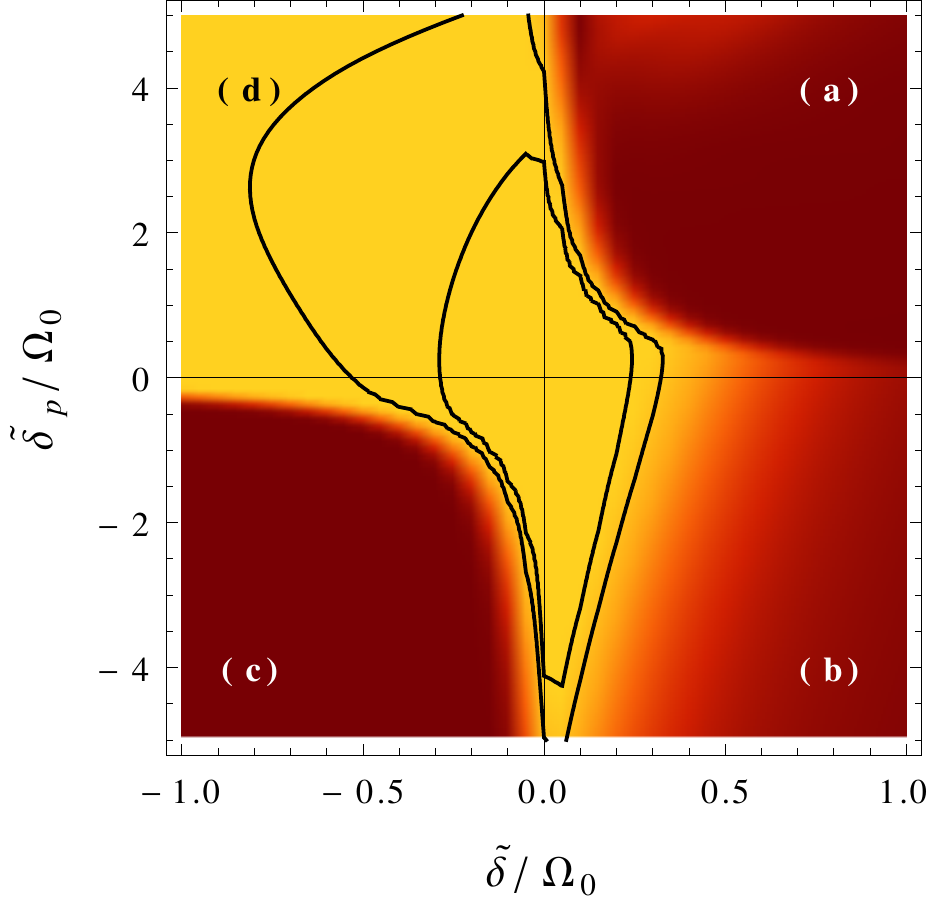}
\caption{(color online) The color map describes the efficiency 
of ideal cSTIRAP, with $\Gamma_2 = 0$ vs fluctuations of the 
detunings. In the brightest area we have  
$P_1(t_f)>0.9$.
Lines refer to $\Gamma_2=1/T$ and 
delimit the $P_1(t_f)>0.9$ (most inner region) 
and the $P_1(t_f)>0.8$ areas.
We use the same parameters as in Fig.~\ref{fig:eigenvalues}(b),
which guarantee that 
in absence of fluctuations,
$\tilde{\delta}_s=\tilde{\delta}_p=\tilde{\delta}=0$, 
adiabaticity of the protocol is strong. The extension of 
the regions of large efficiency determines the single-photon
linewidths (in this case $\Delta\tilde{\delta}_s$) and the
two-photon linewidth $\Delta\tilde{\delta}$. 
\label{fig:sens_detunings}}
\end{figure}
The efficiency of the protocol versus stray detunings 
is shown in Fig.~\ref{fig:sens_detunings}. 
The colour map shows $P_1(t)$ for $\Gamma_2 =0$ 
at the end of the protocol, $t=t_f$. 
Lines refer to finite $\Gamma_2 =1/T$, which determines a  
reduced value of $P_1(t_f)$ since a nonvanishing 
population $P_2(t)$ would decay outside the system.
It is seen that the efficiency is large in a whole region 
around the center of the plot (absence of fluctuations,  
$\tilde{\delta}_s=\tilde{\delta}_p=\tilde{\delta}=0$),
showing the stability of the protocol. The failure 
of cSTIRAP in the region of larger detunings is analyzed in 
the App.~\ref{sec:failure-stirap}. Here we mention that   
in the first and in the third quadrants 
of Fig.~\ref{fig:sens_detunings} failure is due to ``bad 
projection'' errors, i.e. the system may evolve along an 
adiabatic linkage leading to a wrong target state. Instead 
deep in the second quadrant the problem is ``bad adiabaticity''
due to an insufficient pump-induced two-photon 
avoided crossing.

Concerning sensitivity to $\tau$, notice that the 
convenient delay  is implicitly set by the choice of 
$\Omega_p(t)$ being maximal at the second crossing time, Eq.(\ref{eq:pulse}). 
We have checked that in these conditions the protocol is 
stable against deviations from the delay and the 
detailed pulse shape used in this work, provided they are 
not too large. Moreover it is worth stressing 
that the protocol we propose in the ``ideal'' detunings case, 
while being physically satisfactory, is not an 
optimal solution in the mathematical sense. 
Therefore we expect further improvement 
by tackling the problem with Optimal Control Theory.

\section{Decoherence}
\label{sec:decoherence}
A further important source of errors in STIRAP is 
decoherence~\cite{ka:213-falci-prb-stirapcpb}, 
especially in solid-state artificial atoms. 
We discuss some qualitative aspect in this section. 
A key asset of conventional STIRAP is that while 
spontaneous decay from $\ket{2}$ may be large (decay time 
larger than 
the duration of the protocol), the phenomenon is supposed 
to have small impact as long as $\ket{2}$ is depopulated. 
This holds true also for cSTIRAP, as seen from the results for 
$\Gamma_2 \neq 0$ presented in the last Section.
Markovian dephasing in STIRAP has been studied in 
detail~\cite{ka:204-ivanov-pra-stirapdephasing} 
and its detrimental effect, namely leakage from the 
trapped subspace due to the weakening of 
destructive interference phenomenon, has been elucidated. 
It has been shown that strong Markovian dephasing is tolerated, 
as long as it does not affect the two levels of the trapped 
subspace.
More complete studies of the effects of quantum noise in driven 
systems have pointed out that in solid state implementations 
of three-level 
artificial atoms the main effect is due to decay processes 
within the trapped subspace~\cite{ka:213-falci-prb-stirapcpb}.
Other decoherence channels  
emerging in the Born-Markov approximation, namely 
the relation of rates to the detailed spectral density 
of the environment~\cite{ka:195-geva-jorchemphys-gmerabi} 
and the possible drive-induced absorption~\cite{ka:213-falci-prb-stirapcpb}, are less relevant.

On the other hand, 
unlike their natural counterpart, artificial atoms 
implemented by solid-state nanodevices suffer from 
low-frequency noise~\cite{kr:214-paladino-rmp}. 
This drawback may be compensated by  
the ease of producing large couplings on chip, the tradeoff 
between protection and addressability being the central design 
issue. 
The effect of low-frequency noise in STIRAP 
has been discussed in Ref.~\onlinecite{ka:212-falci-physscr}, 
where its interplay with Markovian 
noise and the role of device design were also 
addressed~\cite{ka:213-falci-prb-stirapcpb,ka:215-distefano-brasov}. The 
extension of this detailed analysis to cSTIRAP is beyond the 
scope of this paper, but general features 
pointed out in the above works 
together with the results of the last section, allow to 
draw a physical picture which can be used as a guide for 
device design. 

We assume that low efficiency may be determined by 
by decoherence leading to detrapping from the 
$\{\ket{0},\ket{1}\}$ subspace and by failures of 
the adiabatic approximation also leading to unwanted 
population of $\ket{2}$. The simplest model  
encompassing these main features is to 
account for decay of $\ket{2}$ in a continuum due to 
quantum noise ($\Gamma_2$) and to account for dephasing as 
due to low-frequency (classical) fluctuations of relevant 
parameters. 
That is we consider the Hamiltonian
Eq.(\ref{eq:H}) supplemented by the non-Hermitian term 
appearing in Eq.(\ref{eq:parametric-hamiltonian}), and let
$\delta_k(t) \to \delta_k(t) + \tilde{\delta}_k(t)$, 
for $k = s,p$, and 
$\Omega_k(t) \to \Omega_k(t) + \tilde{\Omega}_k(t)$, 
where $\tilde{\delta}_k(t)$ and $\tilde{\Omega}_k(t)$
are classical stochastic processes. In artificial atoms 
such fluctuations stem physically from noisy 
external bias fields, which induce fluctuations of the energy
splittings of the device (determining $\tilde{\delta}_k$'s) 
and of the operator 
coupling to the field (yielding $\tilde{\Omega}_k$'s).
The efficiency is obtained by averaging over such fluctuations
$P_1(t|\{\tilde{\delta}_k\},\{\tilde{\Omega}_k\})$,  
at the end of the protocol.
In cases of interest, as for $1/f$ noise, the average can be 
estimated in the quasistatic (or static-path) 
approximation~\cite{ka:205-falci-prl,kr:214-paladino-rmp}. 
It amounts to 
substitute stochastic processes by random variables with a 
suitable Gaussian distribution, which physically 
accounts for sample to sample fluctuations of parameters. 
Results of the last section suggest that stray
$\tilde{\Omega}_k$s hardly affect the efficiency, whereas 
the effect of the distribution of  $\tilde{\delta}_k$'s 
can be important. This effect can be 
read off in Fig.\ref{fig:sens_detunings}, which shows that 
for reasonably small fluctuations there is a region where 
still large efficiencies are allowed. 
Successful cSTIRAP requires that fluctuations 
of energy levels (i.e. detunings) are smaller than the 
linewidths. 
In analogy with the analysis 
of Ref~\onlinecite{ka:213-falci-prb-stirapcpb} we expect that
the condition of large efficiency depends on 
the bandstructure of the device at the bias 
point. Indeed depending on the device and on the noise source, 
fluctuations of the two splittings (detunings) are 
either correlated or anticorrelated~\cite{ka:215-distefano-brasov}, namely they are described by 
lines with positive or negative slope in  
Fig.~\ref{fig:sens_detunings}. A figure of merit is the 
ratio $\delta_{1 \over 2}/ \sigma_\delta$ between the 
two-photon linewidth of STIRAP, corresponding to the width 
of the large efficiency region in the proper direction in 
Fig.~\ref{fig:sens_detunings}, and the variance 
$\sigma_\delta$ of the fluctuations of the two-photon detuning.

\section{Conclusions}
\label{sec:conclusions}
In this paper we have proposed a new protocol 
which extends conventional STIRAP.  
Coherent population transfer is achieved 
with reduced available control, namely one of the field is 
kept always on.
This procedure is suited for applications 
in artificial atoms and can be advantageous in integrated 
atom-cavity systems architectures, where couplings to 
quantized modes are implemented by non-switchable 
hardware~\cite{kr:208-schoelkopf-nature-wiring}, 
and may be manipulated in this way for applications to 
microwave quantum photonics~\cite{ka:213-nakamurayam-ieee-microwphot}. In this respect it may be useful that cSTIRAP can be 
repeated cyclically since population histories are invariant    
when $\delta_k \to -\delta_k$, allowing the protocol to work  
as well in the reverted detunings configuration.

The protocol leverages on the fact that in the microwave realm 
external fields have a phase 
which can be usually controlled better 
than for sources at optical frequencies.  In particular 
frequency can be modulated more accurately allowing 
direct time-dependent control of the detunings, instead of the 
induced Stark shifts used in genuine 
atomic systems~\cite{kr:201-vitanov-annurev}. 
Moreover in solid-state  
artificial atoms, e.g. superconductor based, detunings 
can be independently modulated by external voltages 
and fluxes. 

Manipulation of detunings is the basis of other coherent 
transfer protocols like RCAP~\cite{ka:199-solabergmann-pra-rcap}.
The essential difference between standard RCAP and cSTIRAP 
is that, owing to the fact that the  
Stokes field is always-on, our protocol involves a dressed 
state in the AP phase (see Sec.~\ref{sec:cSTIRAP-mech}),  
whereas in the former AP occurs between bare states. 
Therefore while in RCAP the avoided 
crossing is due to the two-photon coupling of two far 
detuned dispersively coupled fields, in cSTIRAP 
AP takes place via 
destructively interfering {\em resonant} fields.
This renders more robust 
the protocol, which achieves large efficiency for rather 
small pulse duration. On the other hand the analogy with RCAP, 
as well as  
the discussion of Sec.~\ref{sec:decoherence},  
suggests that also cSTIRAP may be 
resilient to phase noise and to low-frequency noise in 
nanocircuits offering advantages in quantum state processing
with artificial atoms~\cite{ka:208-weinori-prl-stirapqcomp}. 

STIRAP is also the basis of other protocols as preparation 
of superpositions~\cite{kr:201-vitanov-annurev},
transfer of wavepackets~\cite{kr:207-kral-rmp-controladpass},
manipulation of photons and 
quantum gates~\cite{ka:208-weinori-prl-stirapqcomp},
with still unexplored potentialities for quantum information 
and quantum control. Therefore demonstration of cSTIRAP 
is a benchmark for a class of multilevel advanced control 
protocols in artificial atoms.

\appendix
\section{Adiabatic elimination of state $\ket{2}$}
\label{sec:app-adiabatic}
In order to estimate the population of $\ket{2}$ we 
start from the usual adiabatic elimination in the bare basis. 
The Schr\"odinger equation $i\partial_t\ket{\psi} = H\ket{\psi}$, with the Hamiltonian Eq.(\ref{eq:H}), is written for the 
components of $\ket{\psi}:=\sum_{i=0}^2 c_i \ket{i}$. 
Assuming $\dot{c}_2 \simeq 0$ one finds 
\begin{equation}
\label{eq:c2}
c_2 = - \frac{\Omega_p c_0+\Omega_s c_1}{2\delta_p}
\end{equation}
This expression of $c_2$ is substituted in the Schr\"odinger 
equation yielding a  two-state problem 
governed by the effective Hamiltonian
\begin{equation}
\label{eq:Hae}
H_2(t) = \begin{bmatrix}
-\frac{\Omega^2_p}{4\delta_p}	 								& 	-\frac{\Omega_s\Omega_p}{4\delta_p}								\\
-\frac{\Omega_s\Omega_p}{4\delta_p}	 								& 	\delta	-\frac{\Omega^2_s}{4\delta_p}			 	\\
\end{bmatrix}
\end{equation}
Now assuming the validity of the adiabatic approximation, 
$c_0$ and $c_1$ are approximately given by the 
instantaneous eigenvectors of $H_2(t)$. In particular we consider the state corresponding to the preparation $\ket{\psi(t_i)}=\ket{0}$, and we can estimate $P_2 = |c_2|^2$ from
Eq. (\ref{eq:c2}). The analytic result is shown in 
Fig.~\ref{fig:eigenvalues}(b), thin solid line in the inset, and 
it yields good agreement with the numerical curve. The 
analytic 
expression, though easy attainable, is cumbersome. 
Insight in the parametric dependence can be gained 
by evaluating leakage at $t = t_c$:
\[
P_2(t_c) = \frac{\kappa_{\delta}-1}{\kappa_{\delta}} \frac{(\kappa - \sqrt{\kappa^2 + 4})^2}{4 + (\kappa + \sqrt{\kappa^2 + 4})^2}
\]
which is Eq. (\ref{eq:P2-adel}).
\begin{figure}[t!]
\includegraphics[width=0.9\columnwidth]{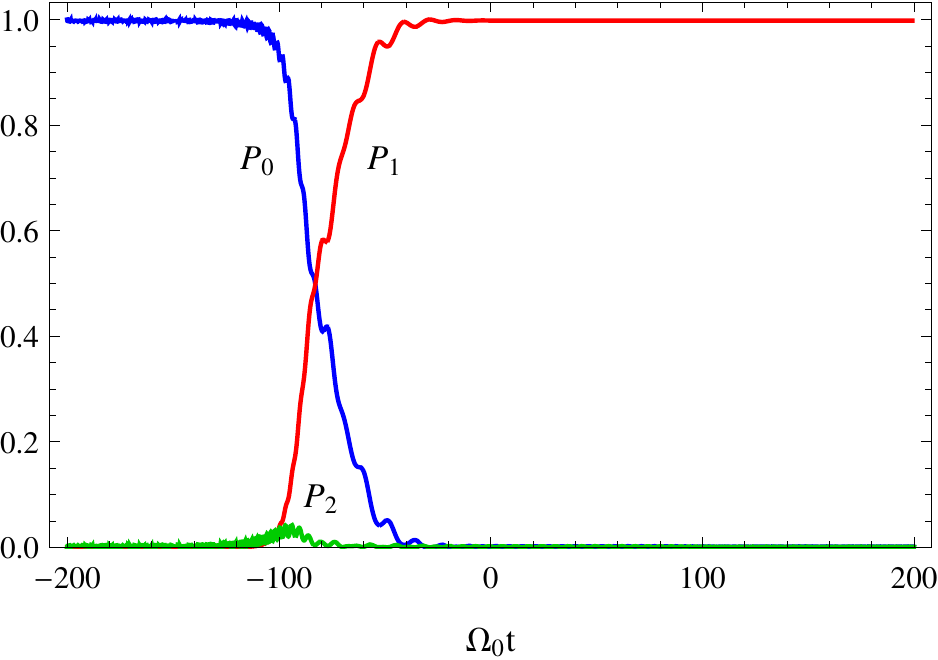}
\caption{(color online) Population histories for $\Omega_p(t) =: \Omega_0, \Omega_0T=40, h_{\delta}=10, \kappa_{\delta}=1.2$, $\kappa = 1$ and $\tau_{ch}=0.6T$.
\label{fig:populationsPAO}}
\end{figure}
We remind that adiabatic elimination yields coarse grained 
amplitudes and it is a priori enforced by large 
single-photon detunings. Remarkably the result obtained from
Eq.(\ref{eq:c2}) is accurate for the whole 
procedure, even if in part of the protocol the 
condition $\delta_p \gg \Omega_k$ is not met. This is 
due to the fact that the population of $\ket{2}$ is always 
small, either because the regime is dispersive or because 
there is destructive interference. 

Corrections in the regime where $\delta_p(t) \lesssim \Omega_p,\Omega_s$ can be fully taken into account if adiabatic elimination
is carried in the representation of the Stokes 
eigenstates. We write the Hamiltonian Eq.(\ref{eq:H}) 
in the basis $\{\ket{0},\ket{s_+},\ket{s_-}\}$, given by 
$\ket{s_{\pm}} = a_1^{\pm}\ket{1} + a_2^{\pm}\ket{2}$. 
By expressing $\ket{\psi}=c_0 \ket{0} + c_+\ket{s_+} + c_-\ket{s_-}$ and assuming $\dot{c}_- \simeq 0$, we obtain $c_- = - ({\Omega_-}/{2s_-}) c_0$, where $\Omega_{\pm} = \Omega_p [1+ 4 ({\delta - s_{\mp}}/{\Omega_0})^2]^{-1/2}$. Substituting the Ansatz for $c_-$ into the Schr\"odinger equation yields an effective $2\times2$ Hamiltonian, which in the $\{\ket{0},\ket{s_+}\}$ basis reads:
\begin{equation}
\label{Hae__stokes}
H_{2s} = \begin{bmatrix}
-\frac{\Omega^2_-}{4s_-}	 								& 	-\frac{\Omega_+}{2}								\\
-\frac{\Omega_+}{2}	 								& 	s_+			 	\\
\end{bmatrix}
\end{equation}
which yields the leakage to $\ket{2}$ in the form
\begin{equation}
\label{eq:P2-adel-stokes}
P_2 \simeq |\frac{\Omega_+}{2s_-}c_0\,a_2^- 
+ c_+\,a_2^+|^2
\end{equation}
As it is seen from Fig.~\ref{fig:eigenvalues}(b) (dashed line) 
the result reproduces the numerical solution, 
but it does not yield a figure of merit as simple as 
Eq.~\ref{eq:P2-adel}.


\section{Always-on pump field}
We can seek for a protocol dual to the always-on Stokes field 
by making the following 
substitutions, 
$t_c \to - t_c$, 
$\delta_p \rightleftarrows \delta_s$, 
$\Omega_p \rightleftarrows \Omega_s$. The population histories 
are shown in Fig.~\ref{fig:populationsPAO} and differ somehow 
from those of Sec.\ref{sec:co-po-tra}. The point is that the 
system is prepared in $\ket{0}$, which in this case is not an 
exact eigenstate of the initial Hamiltonian. As a consequence
Rabi oscillations of small amplitude appear in both $P_0$ and 
$P_2$. They can be substantially reduced by increasing 
the initial value of the pump detuning. Stray population may 
appear in the intermediate state $\ket{2}$ also due to 
adiabatic population transfer at the avoided crossing, and can 
be minimized by 
adjusting parameters as suggested by Eq.~\ref{eq:P2-adel}.

\section{Failure of STIRAP at large detunings}
\label{sec:failure-stirap}  
We now analyze the dynamics in the regions 
of Fig.~\ref{fig:sens_detunings} 
where cSTIRAP fails. As mentioned in Sec.~\ref{sec:cSTIRAP-sens}
when energy levels have infinite lifetime, failure of the 
protocol is due to two kind of errors, namely "bad adiabaticity" 
and "bad projection"~\cite{kr:201-vitanov-advmolopt}. 
While in the former case, the protocol fails because the 
avoided crossing produced by the fields is insufficient to 
guarantee adiabaticity, in the latter case the system is 
projected onto the wrong eigenstate of the Hamiltonian.
Errors mainly occur during the AP near the point at $t=t_c$
where Stokes eigenstates cross. 
An efficient protocol requires for the probabilities 
of Landau-Zener transitions between such states that 
$\gamma_{0\to s_-}\ll 1$ and $1-\gamma_{0\to s_+}\ll 1$, which
is not always met for finite stray detunings. 

\begin{figure}[t!]
\includegraphics[width=0.45\columnwidth]{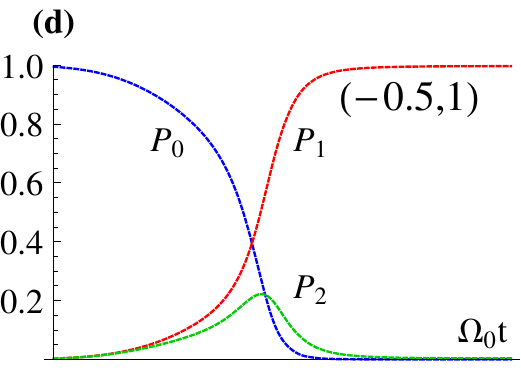}
\hfill
\includegraphics[width=0.45\columnwidth]{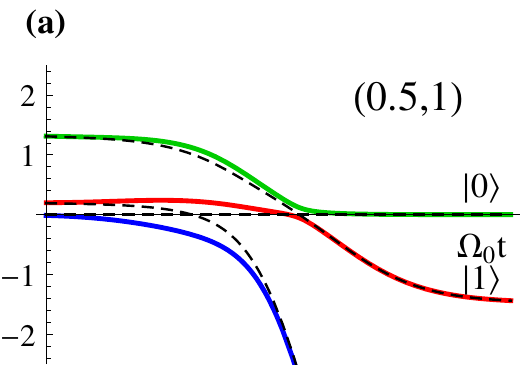}
\\
\includegraphics[width=0.45\columnwidth]{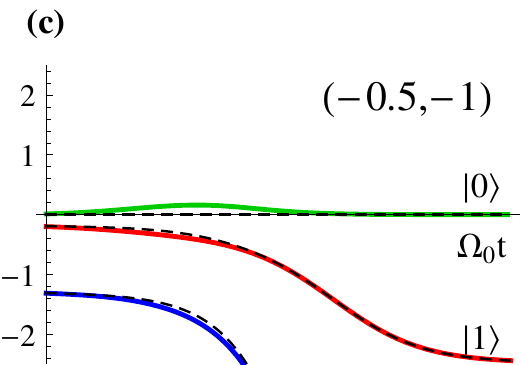}
\hfill
\includegraphics[width=0.45\columnwidth]{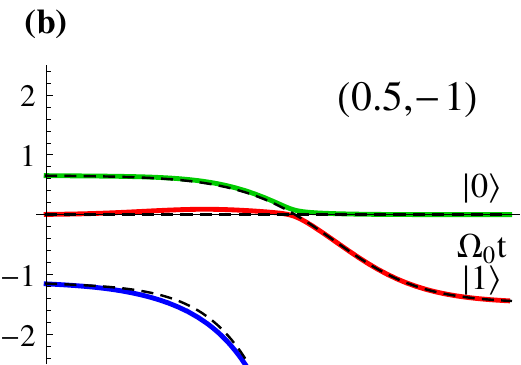}
\caption{(color online) (a-c) Instantaneous eigenvalues 
of the Stokes (dashed lines) and complete (solid lines) 
Hamiltonians, in the dark regions of quadrants (a-c) of 
Fig.~\ref{fig:sens_detunings}, in units of $\Omega_0$.  
(d) Population histories corresponding to  
quadrant (d) of Fig.~\ref{fig:sens_detunings} for $\Gamma_2=0$, 
showing that, even if $P_1(t_f)$ is nearly one, the protocol 
suffers of large transient population of $\ket{2}$ (green line). 
In each panel the label $(\tilde{\delta}/\Omega_0,\tilde{\delta}_p/\Omega_0)$ indicates the value of the stray detunings.
\label{fig:sens_detunings2}}
\end{figure}

A qualitative picture of how cSTIRAP possibly fails due to 
stray detunings is offered by the patterns of the instantaneous 
eigenvalues of the full and of the Stokes Hamiltonians, in 
the darker regions of the three (a-c) quadrants of 
Fig.~\ref{fig:sens_detunings}. 
Examples of these patterns are plotted in  
Fig.~\ref{fig:sens_detunings2}(a-c).

In the region deep in quadrant (a) 
of Fig.~\ref{fig:sens_detunings}
detunings are such that the first crossings of the 
Stokes eigenenergies occurs 
{\em at positive times}, i.e. when $\Omega_p$ is already on
[Fig.~\ref{fig:sens_detunings2}(a)]. 
Therefore $\ket{0}$ and  $\ket{s_-}$ mix, originating 
a sort of initial ``bad projection'' error. Then the subsequent
swap   $\ket{s_-}\to \ket{2}$ leads to a wrong target state. 
Deep in quadrant (b), the protocol suffers from 
a sort of final ``bad projection'' error: 
the second crossing occurs {\em at negative times}, 
where $\Omega_p \approx 0$ and the correspondent 
transition becomes diabatic. This yields 
$\ket{\psi(t)}\approx \ket{0}$ at all times~\cite{note:times}. 
Deep in quadrant (c) cSTIRAP fails when the configuration 
of detunings renders the pump-induced avoided crossing 
insufficient. In this case the problem 
is ``bad adiabaticity'', Zener tunneling inducing 
unwanted transitions to the state adiabatically evolving 
towards $\ket{0}$.

Finally, deep in the quadrant (d) the configuration of 
detunings is such that the two ``mixing'' phases of the 
protocol are inverted. Indeed the Stokes-induced AT splitting 
becomes relevant only {\em after} the second crossing, 
which in the ideal case would have produced the two-photon AP.
Therefore $\Omega_p$ partially injects population into 
$\ket{2}$. At later times, in the Stokes-AT phase, this 
population is swapped to $\ket{1}$. Although the final state is 
correct (cf. the large efficiency in 
Fig.~\ref{fig:sens_detunings}), in the presence of decay $\Gamma_2 \neq 0$, occupation of $\ket{2}$ at intermediate times suppresses the efficiency [see Fig.~\ref{fig:sens_detunings2}(d) and the solid lines 
of Fig.~\ref{fig:sens_detunings}].


\acknowledgments
This work was partially supported by MIUR through Grant. No.
PON02\_00355\_3391233, ``Tecnologie per l'ENERGia e 
l'Efficienza energETICa - ENERGETIC''. A. D'Arrigo acknowledges partial support by Centro Siciliano di Fisica Nucleare e 
Struttura della Materia.


\begin{thebibliography}{33}
\bibitem{note:times} In describing the behaviour in the first and third quadrant we rely on the distinction between negative and positive times. This asymmetry roots from the choice of switching on $\Omega_p(t)$ at slightly positive times.
\end{thebibliography}
 
\end{document}